\documentclass[amsmath,reprint,prb,longbibliography,bibnotes]{revtex4-2}
\usepackage{amsthm,amssymb,amsfonts,graphicx,verbatim, xcolor,bm} 
\usepackage{hyperref} 
\usepackage[utf8]{inputenc} 
\usepackage{siunitx} 
\usepackage{braket}

\begin{document}

\title{Topological excitonic insulators in electron bilayers modulated by twisted hBN}
\author{Yongxin Zeng}
\email{yz4788@columbia.edu}
\affiliation{Department of Physics, Columbia University, New York, NY 10027}
\author{Allan H. MacDonald}
\affiliation{Department of Physics, University of Texas at Austin, Austin, TX 78712}
\author{Nemin Wei}
\affiliation{Department of Physics, Yale University, New Haven, CT 06520}

\begin{abstract}
Equilibrium interlayer exciton condensation is common in bilayer quantum Hall 
systems and is characterized by spontaneous phase coherence between isolated
layers.  It has been predicted that similar physics can occur in the 
absence of a magnetic field in some two-dimensional semiconductor bilayers. 
In this work we consider the case of two transition metal dichalcogenide (TMD) monolayers 
separated by a twisted hexagonal boron nitride (hBN) bilayer or multilayer. The hBN 
layers suppress tunneling between the TMD layers so that 
phase coherence is spontaneous when it is present.  When twisted, the hBN layers
also form a ferroelectric moir\'e pattern that applies opposite 
triangular-lattice modulation potentials to the two TMD layers. 
We show via mean-field theory that at total 
hole filling per moir\'e unit cell $\nu=1$,
this geometry can favor a chiral p-wave exciton condensate state
in which the quantum anomalous Hall effect 
coexists with counter-flow superfluidity. 
We present a mean-field phase diagram for TMD hole bilayers modulated by 
twisted hBN, discuss the conditions needed for the realization of the p-wave condensate
state, and propose experiments that could confirm its presence.
\end{abstract}

\maketitle

{\it Introduction.}---Bilayer two-dimensional electron systems have a layer degree-of-freedom that supports some of the most spectacular recent discoveries in condensed matter physics. For example, in twisted transiton metal dichalcogenide (TMD) homobilayers,
a delicate interplay between the moir\'e modulation of potential energy and interlayer tunneling leads \cite{wu2019topological, morales2024magic, shi2024adiabatic, devakul2021magic, zhang2025experimental, thompson2025microscopic} 
to layer-spinor Berry phases and associated effective magnetic fields
that underlie the fractional quantum anomalous Hall effect
\cite{cai2023signatures, park2023observation, zeng2023thermodynamic, xu2023observation}. Our interest here is instead in the case in which interlayer tunneling has been suppressed
by a hexagonal boron nitride (hBN) barrier so that
the number of electrons is separately conserved in each layer.
In these electrically isolated bilayers, interlayer Coulomb interactions can lead to 
exciton condensation in which 
$\langle \psi_t^{\dagger} \psi_b\rangle$ is spontaneously non-zero, breaking 
the layer-U(1) symmetry and establishing spontaneous interlayer phase coherence \footnote{We restrict our attention here to the case of equilibrium exciton condensates in which $\langle \psi_t^{\dagger} \psi_b\rangle$ is time-independent.    Related physics occurs in quasi-equilibrium \cite{zeng2020electrically} or non-equilibrium \cite{zeng2024keldysh, sun2024dynamical} exciton condensates in which a bias voltage is applied between layers and $\langle \psi^{\dagger}_t \psi_b\rangle$ has a time-dependent phase.}.
Here $\psi_t^{\dagger}$ and $\psi_b$ are electron creation and annihilation operators in the top ($t$) and bottom ($b$) layers. 
Equilibrium exciton condensation has been established and studied in depth
in bilayer quantum Hall systems \cite{eisenstein2004bose, eisenstein2014exciton, spielman2000resonantly, kellogg2004vanishing, tutuc2004counterflow, kellogg2002drag, li2017excitonic, liu2017quantum, qi2025competition, nguyen2025quantum, zou2024electrical, zou2025vortex}, but its presence in the absence of a magnetic field has not been experimentally confirmed despite recent progress in TMD-based electron-hole bilayers \cite{zeng2020electrically, ma2021strongly, qi2023thermodynamic, qi2025perfect, nguyen2025perfect}.

Moir\'e materials host narrow minibands in a Brillouin zone that is small 
due to the large moir\'e length scale.  Moiré minibands 
can in principle support equilibrium exciton condensates similar to those discovered in quantum Hall systems.  
Previous theoretical work \cite{zeng2022layer, zhang2021su4, Del2024field} has considered TMD
double-bilayers whose moir\'e structures have perfect lateral alignment,
which can be modeled in the strong modulation limit as a lattice of layer pseudospins with antiferromagnetic interactions. Interlayer coherent states are theoretically 
expected in these systems, but the predictions have not been confirmed because of the
difficulty of achieving lateral alignment (see Refs.~\cite{zhang2022correlated, gu2022dipolar, chen2022excitonic, zeng2023exciton} for closely related experiments).

\begin{figure}
    \centering
    \includegraphics[width=\linewidth]{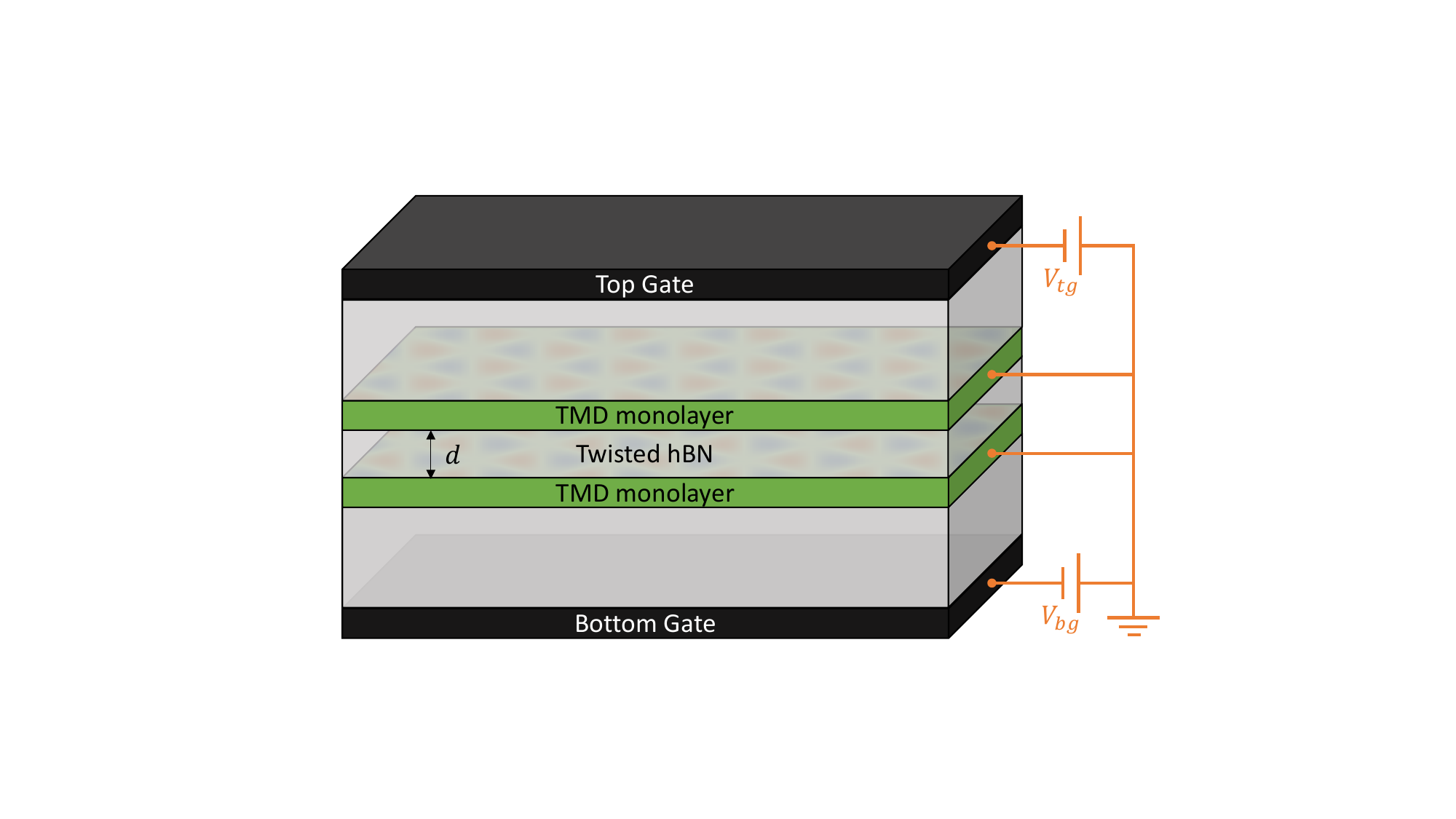}
    \caption{Schematic illustration of the device structure. The twisted hBN layers suppress interlayer tunneling between enclosing top and bottom TMD layers and impose opposite electrostatic modulation potentials (blue and red shading for potential maxima and minima) in the two layers.  The top and bottom gates are separated from the TMD layers by thick untwisted hBN dielectric barriers, and the gate voltages $V_{tg}$ and $V_{bg}$ independently tune the total carrier density and the potential difference $V_D$ between two TMD layers.}
    \label{fig:device}
\end{figure}

In this Letter we consider a bilayer electron system that consists of two TMD monolayers separated by a few-layer twisted hBN spacer as schematically illustrated in Fig.~\ref{fig:device}. The twisted-hBN spacer not only suppresses interlayer tunneling, but also applies periodic modulation potentials onto the top and bottom TMD layers due to its alternating domains of ferroelectricity \cite{yasuda2021stacking, vizner2021interfacial}. This remote imprinting of moir\'e superlattices has 
already proved to be an effective new method to engineer electronic band structure and correlations \cite{kim2024electrostatic, zhang2024engineering, he2024dynamically, gu2024remote, wang2025moire, ghorashi2023topological, ghorashi2023multilayer, zeng2024gate, tan2024designing}. Crucially, the out-of-plane electric polarization of the ferroelectric hBN induces potential energies on the top and bottom TMD layers that are automatically aligned; no lattice alignment is needed between the TMD layers or between the TMD and hBN layers.  When both layers are doped with holes, the carriers are localized near potential maxima that form a triangular lattice, but the triangular lattices in two layers are shifted with respect to each other, forming a honeycomb lattice with layer-sublattice locking. 
We show that at total filling $\nu=1$ per unit cell, the system forms an interlayer coherent state over a wide parameter range. Most interestingly, near zero displacement field and at intermediate moir\'e modulation strength, the ground state is a quantum anomalous Hall (QAH) insulator with {\it spontaneous} chiral p-wave interlayer coherence. 

\begin{figure}
    \centering
    \includegraphics[width=0.7\linewidth]{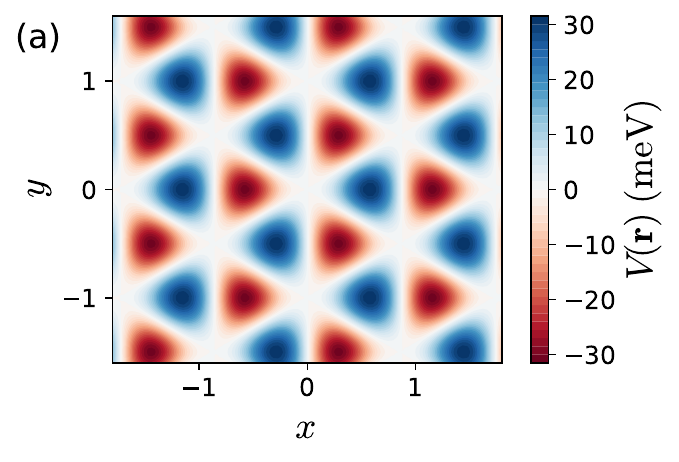}
    \includegraphics[width=\linewidth]{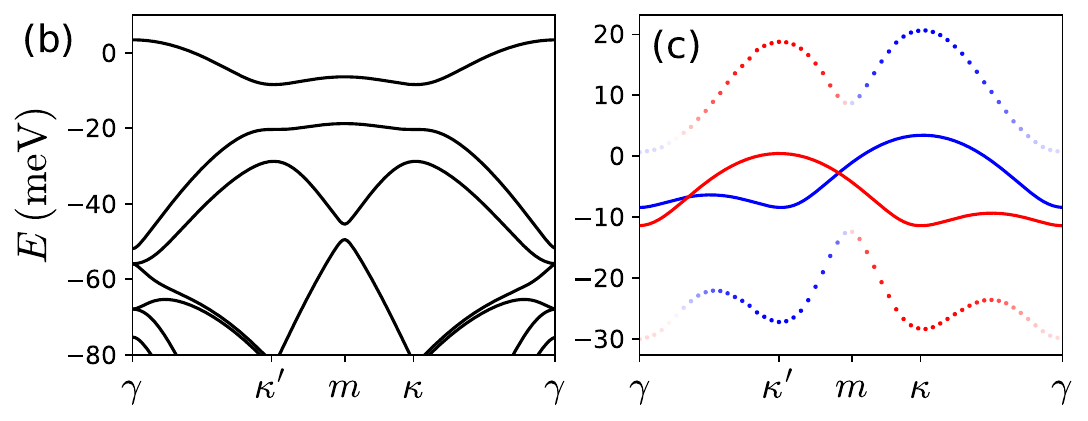}
    \caption{(a) Moir\'e modulation potential $V(\bm r)$ (Eq.~\ref{eq:Vmoire}) with $V_m = \SI{6}{meV}$ and $\psi=\pi/2$ as a model of modulation potential from an adjacent twisted hBN. Lengths are in units of the moir\'e lattice constant $a_m$. 
    Holes in the top layer are localized near the potential maxima shown in blue, while the holes in the bottom layer are localized in the red regions. (b) Band structure of a monolayer TMD with effective mass $m = 0.5\, m_e$ modulated by the potential in (a) with moir\'e length $a_m = \SI{8}{nm}$. (c) The topmost minibands (solid curves) of the top (blue) and bottom (red) TMD layers in a symmetric modulation potential ($\alpha=1$) with displacement field $V_D = \SI{3}{meV}$.  When the two bands are shifted in momentum space by $\bm Q = \bm{\kappa}$ (see main text), the maximum of one band aligns with the minimum of the other, allowing coherence to reduce the Fermi-level density-of-states to zero. The dotted curves show the Hartree-Fock band structure of the p-wave excitonic insulator state at $d=\SI{4}{nm}$ and $\epsilon=6$. The color scale represents layer polarization and shows that the states at $\kappa$ and $\kappa'$ are polarized to opposite layers.}
    \label{fig:bands}
\end{figure}

{\it Model.}---The low-energy physics of a hole-doped TMD monolayer in the presence of a moir\'e modulation potential is described by the continuum model \cite{wu2018hubbard}
\begin{equation}
    H_t = -\frac{\hbar^2 k^2}{2m} + V(\bm r),
\end{equation}
where $m$ is the effective mass of holes in the monolayer TMD and $V(\bm r)$ is the moir\'e modulation potential. In our calculations we take $m = 0.5\, m_e$ where $m_e$ is the free electron mass. Within the first-harmonic approximation, a periodic potential with triangular-lattice periodicity is expressed as
\begin{equation} \label{eq:Vmoire}
    V(\bm r) = 2V_m \sum_{i=1}^3 \cos(\bm g_i \cdot \bm r + \psi),
\end{equation}
where $\bm g_1 = (4\pi/\sqrt{3}a_m,0)$ and its $C_3$-partners $\bm g_2, \bm g_3$ are reciprocal lattice vectors.  In Eq.~\eqref{eq:Vmoire} $V_m$ controls the strength of the modulation potential, and $\psi$ describes its shape, e.g., the ratios of potential values at high-symmetry points. For twisted hBN, the symmetry of the domain pattern
implies that $\psi = \pi/2$ (see Fig.~\ref{fig:bands}(a)). 
The moir\'e lattice constant $a_m = a_{\rm BN}/\theta$ where $a_{\rm BN}$ is the microscopic lattice constant of hBN and $\theta$ is the twist angle.  In general
the top TMD layer is described by single-particle Hamiltonain $H_t$ with potential $V_t$, and the bottom layer by Hamiltonian $H_b$ with potential $V_b$.  
We consider the general case of $N$-layer hBN twisted on top of $M$-layer hBN. 
This allows us to tune the interlayer distance $d$ and the moir\'e potential strength $V_m$, as well as the ratio of the moir\'e potential strengths in two layers $\alpha$. 
The bottom layer is described by the Hamiltonian
\begin{equation}
    H_b = -\frac{\hbar^2 k^2}{2m} - \alpha V_t(\bm r) - V_D,
\end{equation}
where $V_D$ is the potential difference between two layers due to the perpendicular displacement field controlled by the top and bottom gate voltages. 
Without loss of generality, we assume $M\ge N$ so that $0<\alpha\le 1$. Electron-electron interactions are described by the Hamiltonian
\begin{equation}
    H_{\rm int} = \frac{1}{2A} \sum_{ll'} \sum_{\bm{kk}'\bm q} V_{ll'}(\bm q) c_{\bm k + \bm q, l}^{\dagger} c_{\bm k' - \bm q, l'}^{\dagger} c_{\bm k', l'} c_{\bm k, l},
\end{equation}
where $l=t,b$ is the layer index, $c_{\bm k, l}^{\dagger}$ and $c_{\bm k, l}$ are 
momentum-space creation and annihilation operators for electrons in layer $l$, and $A$ is the area of the two-dimensional system. 
The intralayer and interlayer Coulomb interaction potentials are
\begin{equation} \label{eq:V_q}
V_{ll'}(\bm q) =
\begin{cases}
2\pi e^2/\epsilon q, & l=l', \\
(2\pi e^2/\epsilon q) e^{-qd}, & l\ne l',
\end{cases}
\end{equation}
where $\epsilon$ is the dielectric constant. In our numerical calculations we use 
dual-gate-screened interaction potentials (see the Supplemental Material (SM) \cite{SM}) with gate distance $d_g = \SI{50}{nm}$. As in our previous work \cite{zeng2022layer}, we neglect the spin (valley) degree of freedom, anticipating that the energy scale of spin order is much smaller than that of layer order \footnote{For laterally aligned double-moir\'es described by a Hubbard-like model \cite{zeng2022layer}, the energy scale of spin order is $\sim t^2/U$ with hopping $t$ and intralayer onsite repulsion $U$, while the energy scale of layer order is $\sim t^2/V$ where $V \ll U$ is the interlayer onsite repulsion. As the two moir\'es are laterally shifted, $V$ decreases while $U$ stays unchanged, enlarging the separation of energy scales.}. 
When the spin degree of freedom is restored, we expect a competition between spontaneously spin-polarized states like those studied here and competing states with spin-independent exciton order \cite{wu2015theory}, which does not lead to qualitative changes in our excitonic phase diagrams.

{\it Mean-field theory.}---The single-particle band structure of a monolayer TMD modulated by a periodic potential with $V_m = \SI{6}{meV}$ and $a_m = \SI{8}{nm}$ is shown in Fig.~\ref{fig:bands}(b). Assuming that interactions do not lead to significant band mixing, we keep only one band in each layer and project the interaction Hamiltonian onto the two-band subspace. 
An explicit expression for the projected Hamiltonian is provided in the SM \cite{SM}.

To obtain the phase diagram, we perform self-consistent Hartree-Fock calculations at total filling $\nu=1$ (one hole per unit cell); the Hartree-Fock equations are listed in the SM \cite{SM}. Within Hartree-Fock mean-field theory, an interlayer coherent state is described by the Slater determinant 
\begin{equation}
    \ket{\Psi} = \prod_{\bm k \in {\rm mBZ}} (\cos\frac{\theta_{\bm k}}{2} a_{\bm k - \bm Q, t}^{\dagger} + e^{i\phi_{\bm k}} \sin\frac{\theta_{\bm k}}{2} a_{\bm k + \bm Q, b}^{\dagger}) \ket{0},
\end{equation}
where $a_{\bm k, l}^{\dagger}$ is the creation operator for the Bloch state with momentum $\bm k$ in the top miniband in layer $l$, $\bm k$ is defined within the mini-Brillouin zone (mBZ) of the long-period moir\'e superlattice, $\ket{0}$ is the vacuum state in which both bands are empty, and $2\bm Q$ is the excitonic pairing momentum. We find that the ground state energy is minimized when $\bm Q = \pm \bm{\kappa} = \pm \frac{2\pi}{3a_m}(\sqrt{3},1)$, i.e., when the energy maximum of one layer is paired with one of the energy minima of the other layer, breaking time-reversal symmetry. Below we focus on the $\bm Q = \bm{\kappa}$ pairing state (the $\bm Q = -\bm{\kappa}$ state is its time-reversal partner) and, for convenience, relabel the momentum of the Bloch states in two layers by $\bm k \to \bm k \pm \bm Q$ so that pairing occurs at zero momentum:
\begin{equation} \label{eq:Psi_ILC}
    \ket{\Psi} = \prod_{\bm k \in {\rm mBZ}} (\cos\frac{\theta_{\bm k}}{2} a_{\bm k, t}^{\dagger} + e^{i\phi_{\bm k}} \sin\frac{\theta_{\bm k}}{2} a_{\bm k, b}^{\dagger}) \ket{0}.
\end{equation}
The momentum-shifted band structure in the projected subspace is shown in Fig.~\ref{fig:bands}(c).  The Slater-determinant state \eqref{eq:Psi_ILC} can be viewed as occupying 
${\bm k}$-dependent layer pseudospins with orientations
$\bm{n_k} \equiv (\sin\theta_{\bm k}\cos\phi_{\bm k}, \sin\theta_{\bm k}\sin\phi_{\bm k}, \cos\theta_{\bm k})$. Momentum-space winding numbers of the layer 
pseudospins (with proper phase fixing of the layer-basis states; see SM \cite{SM}) define the pairing symmetry and topology of the interlayer coherent state.

To gain some insight into the energetically favored
pairing symmetry, we note that the potential maxima in each layer form a triangular lattice, but the triangular lattices in the two layers are laterally shifted and together form a honeycomb lattice. For the modulation potential illustrated
in Fig.~\ref{fig:bands}(a), for example, holes in the top and bottom layers are predominantly localized at the blue and red regions, respectively. Because of this layer-sublattice locking, the dominant self-energies generated by 
interlayer interactions correspond to inter-sublattice
near-neighbor hopping on the honeycomb lattice. The mean-field Hamiltonian
therefore takes the form
\begin{equation}
\label{eq:hmf}
    H_{\Delta} \approx -\sum_{\braket{ij}} \Delta_{ij} a_i^{\dagger} a_j,
\end{equation}
where $i,j$ are site labels and $\braket{\dots}$ restricts the sum to nearest neighbors on the honeycomb lattice. 
If the coherence amplitudes are equal for all nearest-neighbor pairs ($|\Delta_{ij}| = \Delta$), $H_{\Delta}$ is formally identical to the familiar 
honeycomb-latttice tight-binding model of graphene. 
The interlayer coherence order parameter has 
$p_x\pm ip_y$ vortices at two of the three high-symmetry points $\gamma,\kappa$, and $\kappa'$ in the mBZ, depending on the relative phase 
of $\Delta_{ij}$ for $\braket{i,j}$ pairs along different directions. If the layer 
polarizations at the two vortices are opposite, the layer pseudospin $\bm{n_k}$ forms a topologically 
nontrivial skyrmion texture in the mBZ, implying that the state is
a QAH insulator state with a non-zero Chern number $|C|=1$.
If interactions are strong, some of the lattice symmetries such as $C_3$-rotational symmetry may be spontaneously broken.   For example, coherence is sometimes
formed only between horizontal pairs of sublattice sites, breaking $C_3$-symmetry
and forming an excitonic dimer state that is topologically trivial. 
We therefore expect that the topologically nontrivial p-wave interlayer coherent state forms at intermediate strength of interactions.

\begin{figure}
    \centering
    \includegraphics[width=0.9\linewidth]{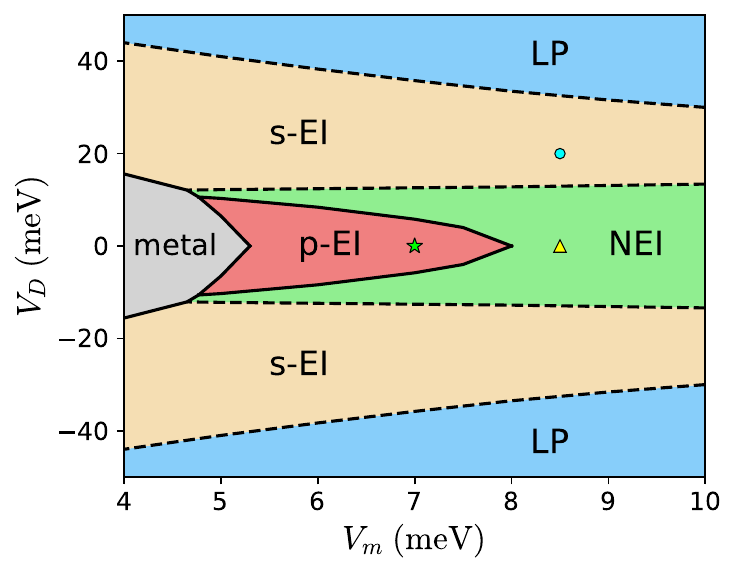}
    \caption{Mean-field phase diagram in the space of modulation potential strength $V_m$ and displacement field $V_D$, with symmetric twisted-hBN modulation ($\alpha=1$), interlayer distance $d=\SI{4}{nm}$, moir\'e length $a_m = \SI{8}{nm}$, and dielectric constant $\epsilon=6$. LP: layer-polarized state; s-EI: s-wave excitonic insulator; p-EI: p-wave excitonic insulator; NEI: nematic excitonic insulator. The solid and dashed lines represent first-order and continuous phase transitions, respectively. The green star, yellow triangle, and cyan circle indicate the parameters for the Berry curvature and layer pseudospin plots in Fig.~\ref{fig:berry_curvature}.}
    \label{fig:phase_diagram}
\end{figure}

\begin{figure*}
    \centering
    \includegraphics[width=0.9\linewidth]{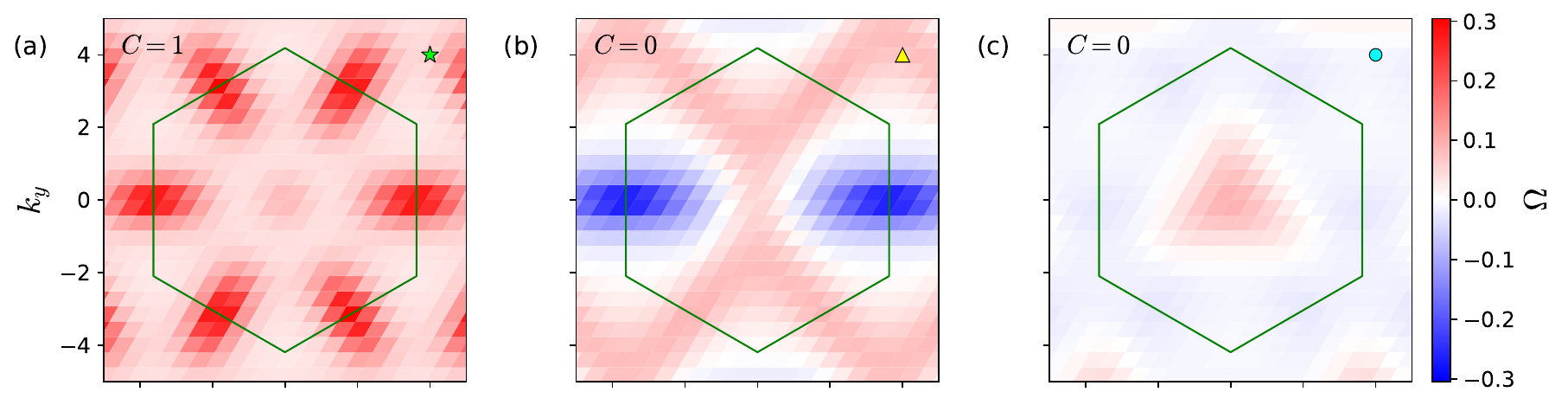}
    \includegraphics[width=0.9\linewidth]{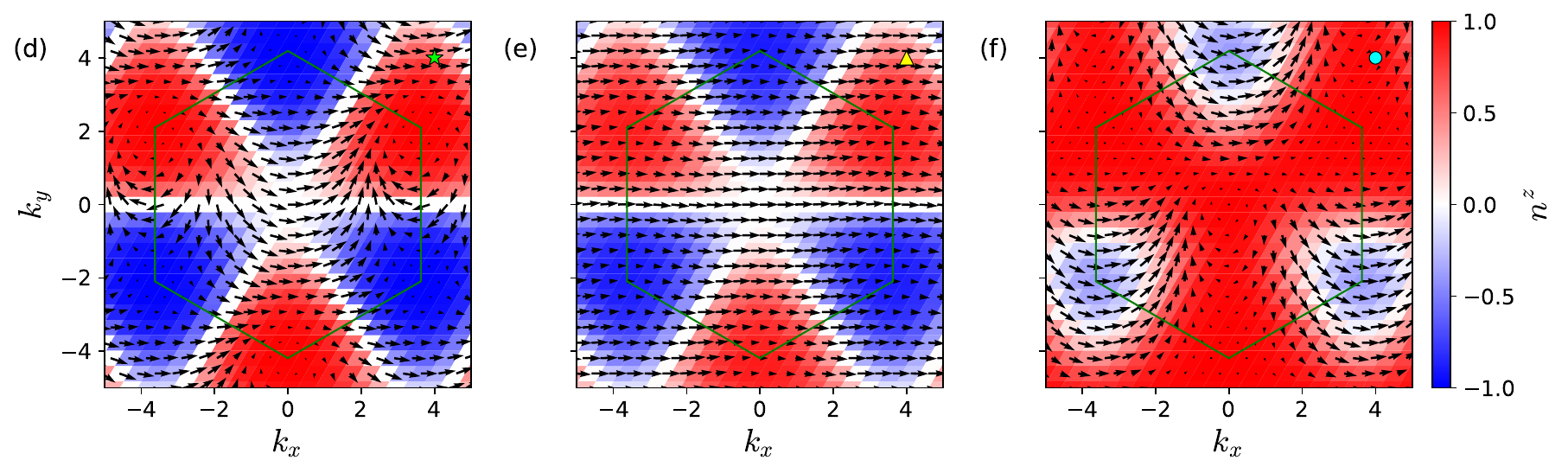}
    \caption{(a-c) Berry curvature and (d-f) layer pseudospin distributions of (a,d) the p-EI state at $V_D = 0, V_m = \SI{7}{meV}$, (b,e) the NEI state at $V_D = 0, V_m = \SI{8.5}{meV}$, and (c,f) the s-EI state at $V_D = 20, V_m = \SI{8.5}{meV}$. The parameters are indicated by the green star, yellow triangle, and cyan circle in Fig.~\ref{fig:phase_diagram}. The $(x,y)$-components of the pseudospins are represented by the black arrows and the $z$-component is shown by the color scale. The unit of length is the moir\'e length $a_m = \SI{8}{nm}$.}
    \label{fig:berry_curvature}
\end{figure*}

{\it Phase diagram.}---Fig.~\ref{fig:phase_diagram} presents the mean-field phase diagram in the space of modulation potential strength $V_m$ and displacement field $V_D$. The remaining parameters are set as $\alpha=1$, $d=\SI{4}{nm}$, $a_m = \SI{8}{nm}$, and $\epsilon=6$; phase diagrams at other values of $\alpha, d, a_m$, and $\epsilon$ are qualitatively similar (see SM \cite{SM}). At large displacement field $V_D$, the system forms a layer-polarized (LP) state, i.e., one of the layers is fully occupied and the other layer is empty. As $V_D$ decreases and the gap between two bands is reduced below a critical value, an excitonic instability occurs and interlayer coherence is established via a continuous phase transition. Just below the critical value of $V_D$, the system is in the dilute boson regime and the ground state is the topologically trivial s-wave excitonic insulator (s-EI).  This state is 
expected because the lowest-energy state of an electron-hole pair is always the 1s-exciton. We find that the topologically nontrivial p-wave excitonic insulator (p-EI) state occurs in a region near $V_D = 0$ and at relatively weak modulation potential strength $V_m$. Between the s-EI and p-EI states is a $C_3$-breaking nematic excitonic insulator (NEI) state that extends to the large-$V_m$ regime. At very small $V_m$, the band width is large and the mean-field ground state is a metallic state.

To further investigate the nature of the three distinct interlayer coherent states, we plot in Fig.~\ref{fig:berry_curvature} the Berry curvature $\Omega_{\bm k}$ and layer (sublattice) pseudospin $\bm{n_k}$ of the ground states at three different sets of parameters. Fig.~\ref{fig:berry_curvature}(d) shows that the topological nature of the p-EI state is essentially the same as that of the Haldane model \cite{haldane1988model, zeng2024sublattice}: the electrons at $\kappa$ and $\kappa'$ are fully polarized to different sublattices, and the in-plane components of sublattice pseudospins form vortices of opposite chiralities around these points. The topologically nontrivial skyrmion-like pseudospin texture in the mBZ indicates a QAH state with Chern number $|C|=1$. For the s-EI state in Fig.~\ref{fig:berry_curvature}(f), on the other hand, the two vortices are located at $\gamma$ and $\kappa$ points in the mBZ where the layer polarizations are the same, leading to a topologically trivial state. The large and slowly varying in-plane component of pseudospins near $\kappa'$ indicates s-wave coherence between the electron and hole pockets from different layers.

For the NEI state shown in Fig.~\ref{fig:berry_curvature}(e), the in-plane component of pseudospins is always nonzero and polarized close to a common in-plane direction at all $\bm k$, indicating a topologically trivial $C_3$-breaking interlayer coherent state. Physically, this represents a state in which coherence is established only between neighboring sublattice sites within a unit cell (horizontal pairs of neighboring sites in Fig.~\ref{fig:bands}(a) by our basis choice) but not across different unit cells. The breaking of $C_3$-symmetry is also clear from the Berry curvature plot in Fig.~\ref{fig:berry_curvature}(b).

To understand the competition between the s-EI, p-EI, and NEI states, we note that at the vortex cores of the momentum-space pseudospin texture, the interlayer coherence order parameter vanishes and does not increase the gap size. It is thus energetically favorable to have s-wave-like interlayer coherence at the high-symmetry point where the two layers are close in energy. When the displacement field $V_D$ is small, the single-particle bands of two layers are nearly degenerate at $\gamma$ and far from degenerate at $\kappa$ and $\kappa'$ (see Fig.~\ref{fig:bands}(c)). It is thus favorable to have a pair of vortices located at $\kappa$ and $\kappa'$ and s-wave coherence at $\gamma$, leading to the topologically nontrivial texture in Fig.~\ref{fig:berry_curvature}(d). When $V_D$ is large and positive, the electron and hole pockets at $\kappa'$ are closest in energy, so the ground state has s-wave coherence at $\kappa'$, leading to the topologically trivial texture in Fig.~\ref{fig:berry_curvature}(f). At intermediate $V_D$, or if $V_m$ is large so that the band width is small, it is favorable to have finite interlayer coherence at all $\bm k$, breaking $C_3$-symmetry.

{\it Discussion.}---We have shown that when two TMD layers are modulated by opposite-sign
triangular-lattice moir\'e potentials, spontaneous interlayer coherence can sometimes lead to a topologically nontrivial excitonic insulator state. The physics in this system is similar to that in AB-stacked MoTe$_2$/WSe$_2$ heterobilayers in which the QAH effect was experimentally observed \cite{li2021quantum, zhao2024realization, tao2024valley} despite weak interlayer hybridization. Recent magnetic circular dichroism measurements \cite{tao2024valley} suggest that in this system the carriers in two TMD layers are polarized to opposite valleys, and it was proposed theoretically \cite{dong2023excitonic, xie2024topological} that the QAH state arises as a result of spontaneous p-wave interlayer coherence. Our phase diagram is qualitatively similar to that obtained in  Ref.~\cite{dong2023excitonic} using a lattice model, although spin order is not explicitly considered in our work. In fact, it was argued in Ref.~\cite{dong2023excitonic} that in the small-$V_D$ regime carriers in each layer are spontaneously spin-polarized due to kinetic ferromagnetism, which further supports the validity of our spinless theory. The competition between intravalley and intervalley coherent states involves microscopic details such as trigonal warping and interlayer tunneling that break valley symmetry and is outside the scope of the present work.

Compared with MoTe$_2$/WSe$_2$ heterobilayers, the system we propose (Fig.~\ref{fig:device}) has the advantage that the two TMD layers are spatially separated by the hBN layers, which guarantees essentially perfect interlayer U(1)-symmetry at the single-particle level and allows for counter-flow measurements when separate electrical contacts are 
made to the two TMD layers.  We predict that a quantized Hall drag effect will occur that is similar to the effect observed in the quantum Hall regime \cite{kellogg2002drag, li2017excitonic, liu2017quantum, qi2025competition, nguyen2025quantum}.  Its observation would not only support the excitonic scenario \cite{dong2023excitonic, xie2024topological} for the MoTe$_2$/WSe$_2$ experiments \cite{li2021quantum, zhao2024realization, tao2024valley}, 
but also provide unambiguous evidence for excitonic coherence in equilibrium
semiconductor bilayers at zero magnetic field that has so far been experimentally elusive. 
This and other properties of this exotic type of exciton condensate remain to be explored.

It is known that Hartree-Fock mean-field theories tend to overestimate the stability of symmetry-broken states. The phase boundary between LP and s-EI states at large $V_D$ is however accurate since it involves the formation and condensation of low-density excitons that are accurately described by mean-field theory. We do expect that when quantum fluctuations are taken into account, the metallic state occupies a larger region in the phase diagram, and that the boundary between the p-EI and NEI states moves to larger $V_m$ (stronger interactions) since the transition to the NEI state involves an additional $C_3$-symmetry breaking. Anticipating that mean-field theory strongly overestimates the tendency to break translational symmetry, we have focused on the weak-correlation parameter regime where charge-ordered crystal states are unlikely. 
Because our model system has a natural real-space formulation, it is 
an ideal playground for quantum Monte Carlo methods \cite{yang2024metal, yang2024honeycomb} including those with the use of neural-network quantum states \cite{carleo2017solving, li2024emergent, luo2024simulating, geier2025attention, li2025deep, luo2025solving} that have recently proved to be efficient methods to simulate interacting electrons in moir\'e materials.

{\it Acknowledgments.---}
We thank Naichao Hu for helpful discussion.
Y.Z. acknowledges support from Programmable Quantum Materials, an Energy Frontiers Research Center funded by the U.S. Department of Energy (DOE), Office of Science, Basic Energy Sciences (BES), under award DE-SC0019443.  A. H. M. was supported by a Simons Foundation Targeted Grant under Award No. 896630. 

\bibliography{references}


\onecolumngrid
\newpage
\makeatletter 

\begin{center}
\textbf{\large Supplemental Material for ``\@title ''} \\[10pt]
\end{center}
\vspace{20pt}

\setcounter{figure}{0}
\setcounter{section}{0}
\setcounter{equation}{0}

\renewcommand{\thefigure}{S\@arabic\c@figure}
\renewcommand{\theequation}{S\@arabic\c@equation}
\makeatother


\section{Details of mean-field calculations}
\subsection{Single-band projection}
The single-particle part of the continuum-model Hamiltonian $H_l$ ($l=t,b$) can be diagonalized in plane-wave basis:
\begin{equation}
    H_l = \sum_{\bm k\in{\rm mBZ}} \sum_n \varepsilon_{\bm k, l}^{(n)} a_{\bm k,l,n}^{\dagger} a_{\bm k,l,n}^{}, \quad a_{\bm k,l,n}^{\dagger} = \sum_{\bm g} u_{\bm k + \bm g, l}^{(n)} c_{\bm k + \bm g, l}^{\dagger},
\end{equation}
where $\bm g$ represents reciprocal lattice vectors. Assuming that the topmost miniband is separated from the other bands by a large enough gap, we keep only the topmost band ($n=0$) from each layer and suppress the band index in the following. The projected Hamiltonian $H = H_0 + H_{\rm int}$ consists of the single-particle term $H_0 = \sum_{\bm k,l} \varepsilon_{\bm k,l} a_{\bm k,l}^{\dagger} a_{\bm k,l}$ and the interaction term
\begin{equation}
    H_{\rm int} = \frac{1}{2A} \sum_{ll'} \sum_{\bm k \bm k' \bm q} V_{ll'}(\bm q) \Lambda_{\bm k}^l(\bm q) \Lambda_{\bm k'}^{l'}(-\bm q) a_{\bm k+\bm q,l}^{\dagger} a_{\bm k'-\bm q,l'}^{\dagger} a_{\bm k',l'} a_{\bm k,l},
\end{equation}
where the form factor $\Lambda_{\bm k}^l(\bm q) \equiv \sum_{\bm g} u_{\bm k+\bm q+\bm g,l}^* u_{\bm k+\bm g,l}$. Here and below $\bm k,\bm k'$ represent momenta within the first mBZ and $\bm q$ runs over the infinite momentum space.

The operator $a_{\bm k,l}^{\dagger}$ has a gauge degree of freedom related to the U(1)-phase choice at each $\bm k$. In order to have a smoothly defined layer pseudospin, we choose a smooth gauge such that the operator $\sum_{\bm k} a_{\bm k,l}^{\dagger}$ creates a localized state in real space. While the gauge for maximally localized Wannier functions can be obtained by the method introduced in Refs.~\cite{marzari1997maximally, marzari2012maximally}, here it suffices to construct a Gaussian wave function $\ket{\psi_{\bm r_0}} = \sum_{\bm{kg}} e^{-(\bm k+\bm g)^2 w^2/2} e^{-i(\bm k+\bm g)\cdot \bm r_0} \ket{\bm k+\bm g}$ with localization length $w$ and localization center $\bm r_0$ and fix the phase of any Bloch state $\ket{u_{\bm k}}$ by the projection
\begin{equation}
    \ket{u_{\bm k}} = \sum_{\bm g} u_{\bm k+\bm g} \ket{\bm k+\bm g} \to \ket{u_{\bm k}} \arg(\braket{u_{\bm k} | \psi_{\bm r_0}}) = \ket{u_{\bm k}} \arg\left[\sum_{\bm g} u_{\bm k+\bm g}^* e^{-(\bm k+\bm g)^2 w^2/2} e^{-i(\bm k+\bm g)\cdot \bm r_0} \right].
\end{equation}
For the moir\'e band states in this work, we choose $\bm r_{0,t} = (a_m/\sqrt{3},0)$ and $\bm r_{0,b} = (2a_m/\sqrt{3},0)$ to be a set of neighboring sites where the moir\'e potential takes its maximum value in the top and bottom layers and $w_t = (\hbar^2 a_m^2/8\pi^2 mV_m)^{1/4}$ and $w_b = w_t \alpha^{-1/4}$ to be the corresponding oscillator lengths near the potential maxima.

\subsection{Hartree-Fock equations}
The Hartree-Fock equations are obtained by decomposing the interaction Hamiltonian $H_{\rm int}$ into the Hartree term
\begin{equation}
    \Sigma_H = \frac{1}{A} \sum_{ll'} \sum_{\bm k \bm k' \bm g} V_{ll'}(\bm g) \Lambda_{\bm k}^l(\bm g) \Lambda_{\bm k'}^{l'}(-\bm g) \rho_{l'l'}(\bm k') a_{\bm k,l}^{\dagger} a_{\bm k,l}
\end{equation}
and the Fock term
\begin{equation}
    \Sigma_F = -\frac{1}{A} \sum_{ll'} \sum_{\bm k \bm k' \bm g} V_{ll'}(\bm k'-\bm k+\bm g) \Lambda_{\bm k}^l(\bm k'-\bm k+\bm g) \Lambda_{\bm k'}^{l'}(\bm k-\bm k'-\bm g) \rho_{l'l}(\bm k') a_{\bm k,l'}^{\dagger} a_{\bm k,l},
\end{equation}
where $\rho_{l'l}(\bm k) \equiv \braket{a_{\bm k,l}^{\dagger} a_{\bm k,l'}} - \delta_{l'l}$ is the single-particle density matrix of the Hartree-Fock ground state subtracted by that of the fully filled valence band. In the numerical calculations we use the dual-gate-screened Coulomb interaction (see Ref.~\cite{kang2020nonabelian} for a derivation)
\begin{align}
V_{ll'}(q) &= \frac{2\pi e^2}{\epsilon q} \frac{(e^{qd}-e^{-qd_g})(e^{-qd}-e^{-qd_g})}{1-e^{-2qd_g}}, \quad l=l', \\
V_{ll'}(q) &= \frac{2\pi e^2}{\epsilon q} \frac{e^{qd}(e^{-qd}-e^{-qd_g})^2}{1-e^{-2qd_g}}, \quad l\ne l',
\end{align}
where $d_g$ is the distance between the top and bottom gates. In our calculations we take $d_g = \SI{50}{nm}$ although the precise value of $d_g$ does not affect our main results. The mean-field ground state is obtained by solving the Hartree-Fock equations self-consistently. The numerical calculations are performed on a $15\times 15$ $\bm k$-grid in the mBZ and with reciprocal lattice vectors cut off at $4|\bm g_1|$ which ensures convergence of the results.

\section{Phase diagrams}
\begin{figure}
    \centering
    \includegraphics[width=0.32\linewidth]{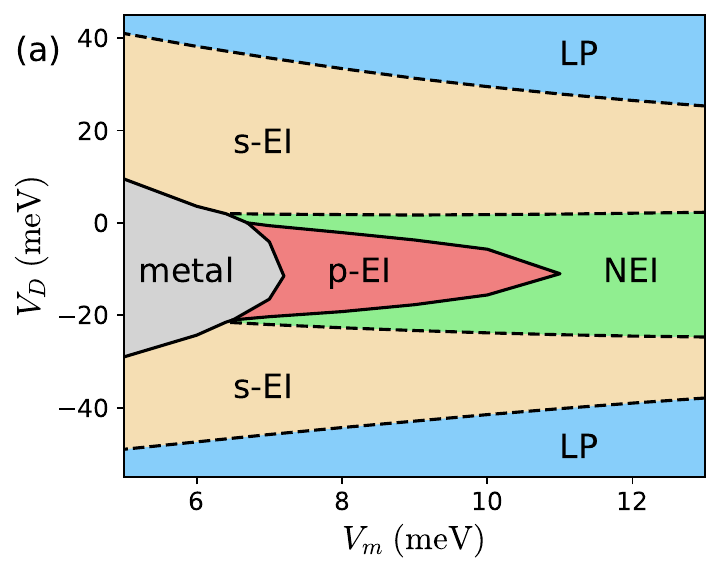}
    \includegraphics[width=0.328\linewidth]{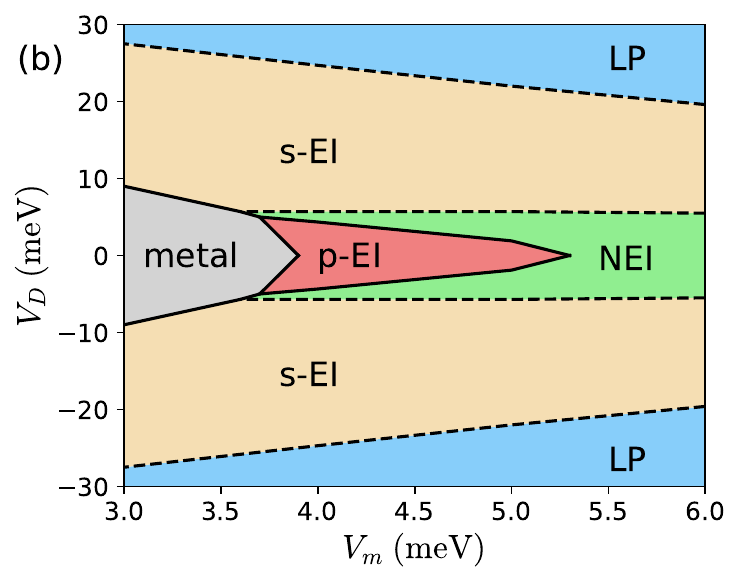}
    \includegraphics[width=0.32\linewidth]{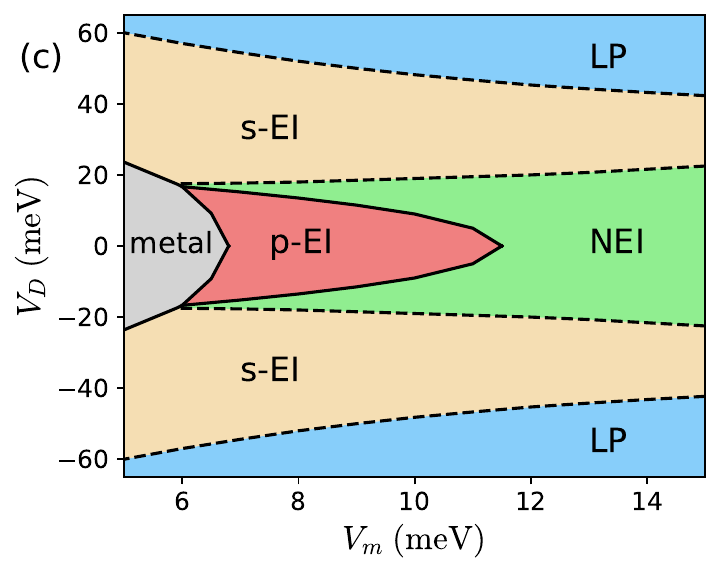}
    \includegraphics[width=0.32\linewidth]{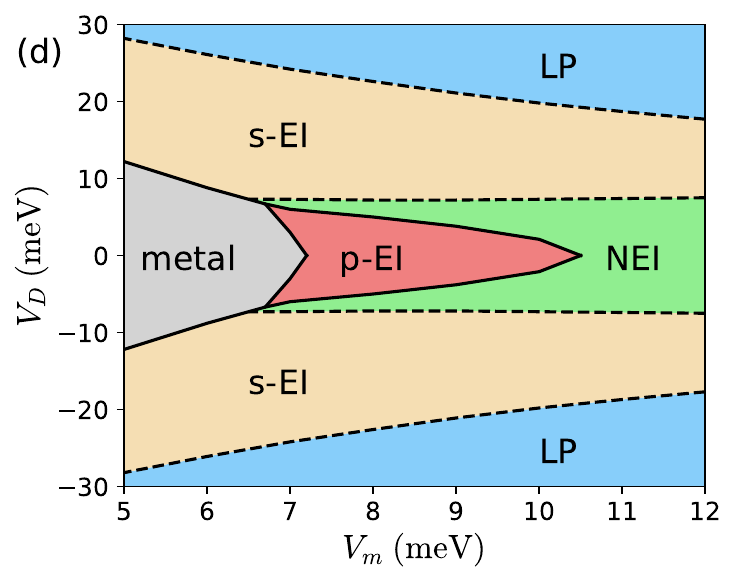}
    \includegraphics[width=0.325\linewidth]{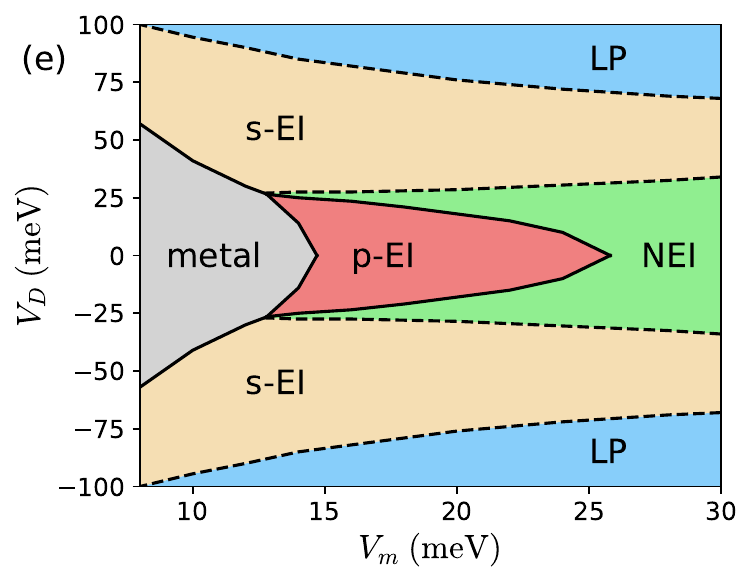}
    \includegraphics[width=0.32\linewidth]{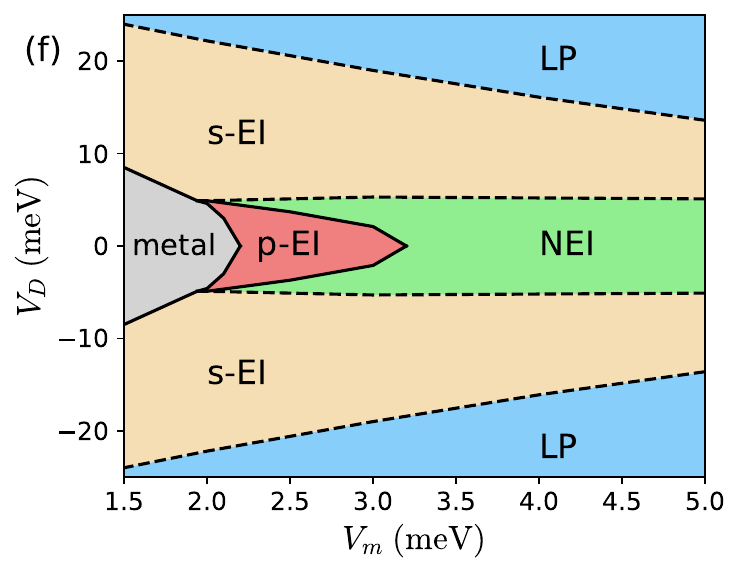}
    \caption{Mean-field phase diagrams at (a) $\alpha=0.5, \epsilon=6, d=\SI{4}{nm}, a_m=\SI{8}{nm}$; (b) $\alpha=1, \epsilon=6, d=\SI{3}{nm}, a_m=\SI{8}{nm}$; (c) $\alpha=1, \epsilon=6, d=\SI{5}{nm}, a_m=\SI{8}{nm}$; (d) $\alpha=1, \epsilon=10, d=\SI{4}{nm}, a_m=\SI{8}{nm}$; (e) $\alpha=1, \epsilon=6, d=\SI{4}{nm}, a_m=\SI{6}{nm}$; (f) $\alpha=1, \epsilon=6, d=\SI{4}{nm}, a_m=\SI{10}{nm}$.}
    \label{fig:phase_sm}
\end{figure}

Fig.~\ref{fig:phase_sm} shows the mean-field phase diagrams at different parameters (layer asymmetry $\alpha$, dielectric constant $\epsilon$, interlayer distance $d$, and moir\'e length $a_m$) from Fig.~\ref{fig:phase_diagram} in the main text. All the phase diagrams are qualitatively similar in trend, but the precise parameter $(V_m,V_D)$ values of phase boundaries are sensitive to the other parameters $(\alpha,\epsilon,d,a_m)$. Roughly speaking, the metallic and p-EI phases tend to occupy a larger region in the phase diagram when the interaction strength is reduced by increasing $\epsilon$, increasing $d$, or decreasing $a_m$.

\end{document}